\documentclass[aps,preprint,showpacs,showkeys,nofootinbib,floatfix,superscriptaddress]{revtex4}
\usepackage{epsfig}
\usepackage{graphicx}
\usepackage{multirow}
\usepackage{tablefootnote}

\newcommand{\br}{\left\langle}
\newcommand{\et}{\right\rangle}

\begin{document}
 
\title{\textbf{Stable bound states of $N$'s, $\Lambda$'s and $\Xi$'s}}
\author{H.~Garcilazo} 
\email{humberto@esfm.ipn.mx} 
\affiliation{Escuela Superior de F\' \i sica y Matem\'aticas, \\ 
Instituto Polit\'ecnico Nacional, Edificio 9, 
07738 M\'exico D.F., Mexico} 
\author{A.~Valcarce} 
\email{valcarce@usal.es} 
\affiliation{Departamento de F\'\i sica Fundamental and IUFFyM,\\ 
Universidad de Salamanca, E-37008 Salamanca, Spain}
\author{J.~Vijande}
\email{javier.vijande@uv.es}
\affiliation{Unidad Mixta de Investigaci\'on en Radiof{\'\i}sica e Instrumentaci\'on Nuclear en Medicina (IRIMED),
Instituto de Investigaci\'on Sanitaria La Fe (IIS-La Fe)-Universitat de Valencia (UV) and IFIC (UV-CSIC), Valencia, Spain}
\date{\today} 

\begin{abstract}
We review our recent work about the stability of strange few-body systems
containing $N$'s, $\Lambda$'s, and $\Xi$'s. 
We make use of local central Yukawa-type Malfliet-Tjon interactions reproducing 
the low-energy parameters and phase shifts of the nucleon-nucleon
system and the latest updates of the hyperon-nucleon and 
hyperon-hyperon ESC08c Nijmegen potentials. 
We solve the three- and four-body bound-state problems by means of Faddeev 
equations and a generalized Gaussian variational method, respectively.
The hypertriton, $\Lambda np$ $(I)J^P=(1/2)1/2^+$, is bound by 144 keV;
the recently discussed $\Lambda nn$ $(I)J^P=(1/2)1/2^+$ system is unbound,
as well as the $\Lambda\Lambda nn$ $(I)J^P=(1)0^+$system, being just above threshold.
Our results indicate that the $\Xi NN$, $\Xi\Xi N$ and $\Xi\Xi NN$ systems 
with maximal isospin might be bound.
\end{abstract}

\pacs{21.45.-v,25.10.+s,11.80.Jy}
\keywords{baryon-baryon interactions, Faddeev equations, variational approaches} 
\maketitle 

\section{Introduction}

Strange nuclear physics is a very topical subject.
The hyperon-nucleon ($YN$) and hyperon-hyperon ($YY$) 
interactions constitute the input for microscopic 
calculations of few- and many-body
systems involving strangeness, such as exotic 
neutron star matter~\cite{Dem10,Ant13,Wei12,Lon14,Mas15} or 
hypernuclei~\cite{Yam10,Hiy08,Yam01}.
There are theoretical debates~\cite{Gar14,Hiy14,Gal14,And15,Afn15,Ric15} 
on the possible existence of a neutral bound state
of two neutrons and a $\Lambda$ hyperon, $^3_\Lambda n$, 
suggested by recent data of the HypHI Collaboration~\cite{Rap13}.
There have been also recent proposals regarding the 
stability of $^4_{\Lambda\Lambda}n$~\cite{Ric15},
the existence of $\Xi$ hypernuclei~\cite{Yam10,Hiy08,Yam01}, or the 
existence of a strangeness $-2$ hypertriton~\cite{Gar13,Gaa16}. 
Obviously, all these predictions are subject
to the uncertainties of our knowledge of the baryon-baryon
interaction, in particular in the strangeness $-2$ sector.
Experimentally, it has been recently reported an emulsion event,
the so-called KISO event, providing evidence of a possible deeply 
bound state of $\Xi^- -^{14}$N~\cite{Naa15}. 
Although microscopic calculations are impossible in this case and, 
consequently, their interpretation will be always affected by
uncertainties, the ESC08c Nijmegen potential has been recently updated~\cite{Nae15,Nag15,Rij16}
to give account for the most recent experimental information of the strangeness
$-2$ sector, the KISO~\cite{Naa15} and the NAGARA~\cite{Tak01} events.
A thorough discussion of the present status of the experimental 
and theoretical progress in hypernuclear physics can be found in Refs.~\cite{Nag10,Gal16}.

When a two-baryon interaction is attractive, if the system is merged with nuclear matter
and the Pauli principle does not impose severe restrictions, the attraction may be reinforced.
Simple examples of the effect of a third or a fourth baryon in two-baryon
systems could be given. The deuteron, $(I)J^P=(0)1^+$, is bound by $2.225$ MeV, while the triton,
$(I)J^P=(1/2)1/2^+$, is bound by $8.480$ MeV, and the $\alpha$ particle, $(I)J^P=(0)0^+$,
is bound by $28.295$ MeV. The binding per nucleon $B/A$ increases as $1:3:7$.
A similar argument could be employed for strangeness $-1$ systems. Whereas the
existence of dibaryon states is still under discussion\footnote{Note that the
pronounced cusp-like structure seen in many $\Lambda N$ related observables
near the $\Sigma N$ threshold could be very well a signature of a dibaryon~\cite{Mas13}.},
the hypertriton $^3_\Lambda$H, $(I)J^P=(0)1/2^+$, is bound with a separation
energy of $130 \pm 50$ keV, and the $^4_\Lambda$H, $(I)J^P=(0)0^+$, is bound
with a separation energy of $2.12 \pm 0.01 \, {\rm (stat)} \, \pm 0.09 \, {\rm (syst)}$ MeV~\cite{Ess15}. 
This cooperative effect of the attraction in 
the two-body subsystems when merged in few-baryon states was also made evident
in the prediction of a $\Sigma NN$ quasibound state in the $(I)J^P = (1)1/2^+$ 
channel very near threshold~\cite{Gar07,Gac07}.
Such $\Sigma NN$ quasibound state has been recently suggested
in $^3\rm{He}(K^-,\pi^\mp)$ reactions at 600 MeV/c~\cite{Har14}.

In this paper, we review our recent studies of the three-body systems:
$\Lambda NN$, $\Xi NN$, $\Lambda\Lambda N$, and $\Xi \Xi N$, as well as the 
four-body systems $\Xi\Xi NN$ and $\Lambda\Lambda NN$. 
We make use of the most recent updates of the
ESC08c Nijmegen potentials in the strangeness $-1$, $-2$, $-3$
and $-4$ sector~\cite{Nag15,Nae15,Rij13} accounting for the
recent KISO~\cite{Naa15} and NAGARA~\cite{Tak01} events in the strangeness 
$-2$ sector. As discussed above, the existence of two-body attractive interactions 
or bound states could give rise to other stable few-body systems when
merged with other nucleons or hyperons. For example, the overall attractive 
character of the $\Xi N$ interaction
comes suggested by recent preliminary results from lattice QCD~\cite{Sas15} 
together with other indications of certain emulsion data~\cite{Rij13,Nag15,Rij16}.
Besides the recent update of ESC08c Nijmegen model, 
$\Xi-$hypernuclear calculations~\cite{Hiy01} and
chiral quark models~\cite{Car12} found a $\Xi N$ attractive
interaction before the KISO event. Furthermore,
the possible existence of stable strange few-body states comes 
reinforced by the attractive character of 
the $\Xi\Xi$ interaction for some partial 
waves~\cite{Bea12,Sto99,Mil06,Hai10,Hai15,Nae15,Rij13}.
It is worth to mention that preliminary studies of the
$\Xi\Xi N$ system~\cite{Bea09} indicate that lattice QCD
calculations of multibaryon systems are now within sight.
Analogously, if a second $\Lambda$ would be added to the uncertain $\Lambda nn$ state, 
the weakly attractive $\Lambda\Lambda$ interaction~\cite{Tak01} and the reinforcement 
of the $\Lambda N$ potential without paying a price for antisymmetry requirements, 
may give rise to a stable bound state~\cite{Ric15}. 

One should bear in mind how delicate is the few-body problem in the regime
of weak binding, as demonstrated in Ref.~\cite{Nem03} for the $^4_{\Lambda\Lambda}$H
system. Besides, there are models for the $YN$ interaction, like the hybrid 
quark--model based analysis of Ref.~\cite{Fuj07}, the effective field 
theory approach of Ref.~\cite{Hai16}, or even some of the
earlier models of the Nijmegen group~\cite{Sto99} that,
in general, predict interactions weakly attractive or repulsive.
One does not expect that these models will give rise to stable three- or four-body
states. However, it is worth to emphasize that current hypernuclei studies~\cite{Yam10,Hiy08,Yam01,Nem03,Hiy01} 
have been performed by means of interactions derived from the 
Nijmegen models and, thus, the present review complements such previous work 
for the simplest systems that can be studied exactly.
To advance in the knowledge of the details of the $YN$ interaction, 
high-resolution spectroscopy of $\Xi$ hypernuclei using $^{12}$C targets in $(K^-,K^+)$ 
reactions has been awaited~\cite{Nak10,Nat10} and it is now planned at J-PARC~\cite{Nax15}.
The new hybrid experiment $E07$ recently
approved at J--PARC is expected to record of the order of $10^4$ $\Xi^-$ stopping 
events~\cite{Nak15}, one order of magnitude larger than the previous $E373$
experiment, and will hopefully clarify the phenomenology of some of 
the systems studied in the present work.

The review is organized as follows. In Sec.~\ref{secII} we describe
the technical details to solve the three-body bound state Faddeev equations as well as
the generalized Gaussian variational method used to look for bound states of 
the four-body problem. In Sec.~\ref{secIV} we construct the two-body amplitudes 
needed for the solution of the bound state three- and four-body problems. The 
results are presented and discussed in Sec.~\ref{secV}.
Finally, in Sec.~\ref{secVI} we summarize our main conclusions.

\section{The three- and four-body bound-state problems}
\label{secII}

In this section we outline the solution of the three- and four-body bound-state problems.
We will restrict ourselves to configurations where all particles are in $S-$wave 
states. The three-body problem has been widely discussed in the literature and
we refer the reader to Refs.~\cite{Gac16,Afn74,Gar90} for a more detailed discussion.
The Faddeev equations for a system with total isospin $I$ and total spin $J$ are,
\begin{eqnarray}
T_{i;IJ}^{i_ij_i}(p_iq_i) = &&\sum_{j\ne i}\sum_{i_jj_j}
h_{ij;IJ}^{i_ij_i;i_jj_j}\frac{1}{2}\int_0^\infty q_j^2dq_j
\int_{-1}^1d{\rm cos}\theta\, 
t_{i;i_ij_i}(p_i,p_i^\prime;E-q_i^2/2\nu_i) 
\nonumber \\ &&
\times\frac{1}{E-p_j^2/2\mu_j-q_j^2/2\nu_j}\;
T_{j;IJ}^{i_jj_j}(p_jq_j) \, , 
\label{eq1} 
\end{eqnarray}
where $t_{i;i_ij_i}$ stands for the two-body amplitudes
with isospin $i_i$ and spin $j_i$. $p_i$
is the momentum of the pair $jk$ (with $ijk$ an even permutation of
$123$) and $q_i$ the momentum of particle $i$ with respect to the pair
$jk$. $\mu_i$ and $\nu_i$ are the corresponding reduced masses,
and $h_{ij;IJ}^{i_ij_i;i_jj_j}$ are spin--isospin coefficients.

Expanding the amplitude $t_{i;i_ij_i}(x_i,x_i^\prime;e)$
in terms of Legendre polynomials, Eq.~(\ref{eq1}) 
can be written as,
\begin{equation}
T_{i;IJ}^{i_ij_i}(x_iq_i) = \sum_n P_n(x_i) T_{i;IJ}^{ni_ij_i}(q_i) \, ,
\label{eq11}
\end{equation}
where $T_{i;IJ}^{ni_ij_i}(q_i)$ satisfies the one-dimensional integral equation,
\begin{equation}
T_{i;IJ}^{ni_ij_i}(q_i) = \sum_{j\ne i}\sum_{mi_jj_j}
\int_0^\infty dq_j A_{ij;IJ}^{ni_ij_i;mi_jj_j}(q_i,q_j;E)\;
T_{j;IJ}^{mi_jj_j}(q_j) \, , 
\label{eq12}
\end{equation}
with
\begin{eqnarray}
A_{ij;IJ}^{ni_ij_i;mi_jj_j}(q_i,q_j;E)= &&
h_{ij;IJ}^{i_ij_i;i_jj_j}\sum_r\tau_{i;i_ij_i}^{nr}(E-q_i^2/2\nu_i)
\frac{q_j^2}{2}
\nonumber \\ &&
\times\int_{-1}^1 d{\rm cos}\theta\;\frac{P_r(x_i^\prime)P_m(x_j)} 
{E-p_j^2/2\mu_j-q_j^2/2\nu_j} \, .
\label{eq13} 
\end{eqnarray}

The four-body problem has been addressed by means of the variational
method, specially suited for studying low-lying states. The nonrelativistic
hamiltonian is be given by,
\begin{equation}
H=\sum_{i=1}^4\left(m_{i}+\frac{\vec p_{i}^{\,2}}{2m_{i}}\right)+\sum_{i<j=1}^4V(\vec r_{ij}) \, ,
\label{ham}
\end{equation}
where the potential $V(\vec r_{ij})$ corresponds to an arbitrary two-body interaction.

The variational wave function must include all possible spin--isospin channels
contributing to a given configuration. For each channel $s$, the wave function will be the tensor product of
a spin ($\left|S_{s_1}\right>$), isospin ($\left|I_{s_2}\right>$), and radial
($\left|R_{s_3}\right>$) component,
\begin{equation}
\label{efr}
\left| \phi _{s}\right>=\left|S_{s_1}\right>\otimes\left|I_{s_2}\right>\otimes\left|R_{s_3}\right> \, ,
\end{equation}
where $s\equiv\{s_1,s_2,s_3\}$. Once the spin and isospin parts are integrated out, 
the coefficients of the radial wave function are
obtained by solving the system of linear equations,
\begin{equation}
\label{funci1g}
\sum_{s'\,s} \sum_{i} \beta_{s_3}^{i} 
\, [\langle R_{s_3'}^{j}|\,H\,|R_{s_3}^{i}
\rangle - E\,\langle
R_{s_3'}^{j}|R_{s_3}^{i}\rangle \delta_{s,s'} ] = 0 
\qquad \qquad \forall \, j\, ,
\end{equation}
where the eigenvalues are obtained by a minimization procedure.

For the description of the four-body wave function 
we consider the Jacobi coordinates:
\begin{eqnarray}
\label{coo}
& &\vec{r}_{NN}=\vec{x} =\vec{r}_{1}-\vec{r}_{2} \, , \nonumber \\
& & \vec{r}_{YY}=\vec{y} =\vec{r}_{3}-\vec{r}_{4} \, , \nonumber \\
& & \vec{r}_{NN-YY} = \vec{z} =\frac{1}{2} \left( \vec{r}_{1} + \vec{r}_{2} \right) -\frac{1}{2} 
\left( \vec{r}_{3}+\vec{r}_{4} \right) \, , \\
& &\vec{R}_{\rm CM} = \vec{R} =\frac{\sum m_{i}\vec{r}_{i}}{\sum m_{i}}\nonumber \, ,
\end{eqnarray}
The total wave function should have well-defined permutation properties under
the exchange of identical particles. The spin part can be written as,
\begin{equation}
\left[(s_1s_2)_{S_{12}}(s_3s_4)_{S_{34}}\right]_{S}\equiv|S_{12}S_{34}\rangle_S \, ,
\end{equation}
where the spin of the two $N$'s ($Y$'s) is coupled to $S_{12}$ ($S_{34}$).
Two identical spin-$1/2$ fermions in a $S=0$ state are antisymmetric $(A)$ under permutations while those coupled to 
$S=1$ are symmetric $(S)$.
We summarize in Table~\ref{spin} the corresponding vectors for each total 
spin together with their symmetry properties\footnote{Being the $N$ and $\Xi$ $I=1/2$ particles,
an analogous table serves for the symmetry properties of the wave function in isospin space.
In the case of the $\Lambda$'s the isospin wave function is symmetric}. 
\begin{table}[t]
\caption{Spin basis vectors for all possible total spin states $(S)$. 
The 'Symmetry' column stands for the symmetry properties of the pair of identical particles.}
\label{spin}
\begin{ruledtabular} 
\begin{tabular}{ccccc}
& $S$ & Vector & Symmetry &\\
\hline
&\multirow{2}{*}{$0$} & $|00\rangle_S$ & AA & \\
&                     & $|11\rangle_S$ & SS & \\
\hline
&\multirow{3}{*}{$1$} & $|01\rangle_S$ & AS & \\
&                     & $|10\rangle_S$ & SA & \\
&                     & $|11\rangle_S$ & SS & \\
\hline
&\multirow{1}{*}{$2$} & $|11\rangle_S$ & SS & \\
\end{tabular}
\end{ruledtabular} 
\end{table}

The most general radial wave function with total orbital angular momentum $L=0$ may 
depend on the six scalar quantities that can be constructed with the Jacobi coordinates of the system, they are: 
$\vec x^{\,2}$, $\vec y^{\,2}$, $\vec z^{\,2}$, $\vec{x}\cdot\vec{y}$, $\vec{x}\cdot\vec{z}$, and $\vec{y}\cdot\vec{z}$. 
We define the variational spatial wave function as a linear combination of {\em generalized Gaussians},
\begin{equation}
\left|R_{s_3}\right>=\sum_{i=1}^{n} \beta_{s_3}^{i} R_{s_3}^i(\vec x,\vec y,\vec z)=\sum_{i=1}^{n} \beta_{s_3}^{i} R_{s_3}^i \, ,
\label{wave}
\end{equation}
where $n$ is the number of Gaussians used for each spin-isospin component. 
$R_{s_3}^i$ depends on six variational parameters:
$a^i_s$, $b^i_s$, $c^i_s$, $d^i_s$, $e^i_s$, and $f^i_s$, one for each scalar quantity. 
Therefore, the four-body system will depend on $6\times n\times n_s$ variational parameters, 
where $n_s$ is the number of different channels allowed by the Pauli principle.
Eq.~(\ref{wave}) should have well-defined permutation symmetry under the exchange of both
$N$'s and $Y$'s,
\begin{eqnarray}
\label{parx}
P_{12}(\vec x	\rightarrow -\vec x)R^i_{s_3}&=&P_xR^i_{s_3}\\ \nonumber
P_{34}(\vec y	\rightarrow -\vec y)R^i_{s_3}&=&P_yR^i_{s_3},
\end{eqnarray}
where $P_x$ and $P_y$ are $-1$ for antisymmetric states, $(A)$, and $+1$ for symmetric ones, $(S)$. 
Thus, one can build the following radial combinations, $(P_xP_y)=(SS)$, $(SA)$, $(AS)$, and $(AA)$:
\begin{eqnarray}
\label{wave2-1}
(SS)\Rightarrow R_1^i&=&
{\rm Exp}\left(-a^i_s\vec x^{\,2}-b^i_s\vec y^{\,2}-c^i_s\vec z^{\,2}-d^i_s\vec x\cdot\vec y-e^i_s\vec x\cdot\vec z-f^i_s\vec y\cdot\vec z\right)\\\nonumber
&+&{\rm Exp}\left(-a^i_s\vec x^{\,2}-b^i_s\vec y^{\,2}-c^i_s\vec z^{\,2}+d^i_s\vec x\cdot\vec y-e^i_s\vec x\cdot\vec z+f^i_s\vec y\cdot\vec z\right)\\\nonumber
&+&{\rm Exp}\left(-a^i_s\vec x^{\,2}-b^i_s\vec y^{\,2}-c^i_s\vec z^{\,2}+d^i_s\vec x\cdot\vec y+e^i_s\vec x\cdot\vec z-f^i_s\vec y\cdot\vec z\right)\\\nonumber
&+&{\rm Exp}\left(-a^i_s\vec x^{\,2}-b^i_s\vec y^{\,2}-c^i_s\vec z^{\,2}-d^i_s\vec x\cdot\vec y+e^i_s\vec x\cdot\vec z+f^i_s\vec y\cdot\vec z\right) \, ,
\end{eqnarray}
\begin{eqnarray}
\label{wave2-2}
(SA)\Rightarrow R_2^i&=&
{\rm Exp}\left(-a^i_s\vec x^{\,2}-b^i_s\vec y^{\,2}-c^i_s\vec z^{\,2}-d^i_s\vec x\cdot\vec y-e^i_s\vec x\cdot\vec z-f^i_s\vec y\cdot\vec z\right)\\\nonumber
&-&{\rm Exp}\left(-a^i_s\vec x^{\,2}-b^i_s\vec y^{\,2}-c^i_s\vec z^{\,2}+d^i_s\vec x\cdot\vec y-e^i_s\vec x\cdot\vec z+f^i_s\vec y\cdot\vec z\right)\\\nonumber
&+&{\rm Exp}\left(-a^i_s\vec x^{\,2}-b^i_s\vec y^{\,2}-c^i_s\vec z^{\,2}+d^i_s\vec x\cdot\vec y+e^i_s\vec x\cdot\vec z-f^i_s\vec y\cdot\vec z\right)\\\nonumber
&-&{\rm Exp}\left(-a^i_s\vec x^{\,2}-b^i_s\vec y^{\,2}-c^i_s\vec z^{\,2}-d^i_s\vec x\cdot\vec y+e^i_s\vec x\cdot\vec z+f^i_s\vec y\cdot\vec z\right) \, ,
\end{eqnarray}
\begin{eqnarray}
\label{wave2-3}
(AS)\Rightarrow R_3^i&=&
{\rm Exp}\left(-a^i_s\vec x^{\,2}-b^i_s\vec y^{\,2}-c^i_s\vec z^{\,2}-d^i_s\vec x\cdot\vec y-e^i_s\vec x\cdot\vec z-f^i_s\vec y\cdot\vec z\right)\\\nonumber
&+&{\rm Exp}\left(-a^i_s\vec x^{\,2}-b^i_s\vec y^{\,2}-c^i_s\vec z^{\,2}+d^i_s\vec x\cdot\vec y-e^i_s\vec x\cdot\vec z+f^i_s\vec y\cdot\vec z\right)\\\nonumber
&-&{\rm Exp}\left(-a^i_s\vec x^{\,2}-b^i_s\vec y^{\,2}-c^i_s\vec z^{\,2}+d^i_s\vec x\cdot\vec y+e^i_s\vec x\cdot\vec z-f^i_s\vec y\cdot\vec z\right)\\\nonumber
&-&{\rm Exp}\left(-a^i_s\vec x^{\,2}-b^i_s\vec y^{\,2}-c^i_s\vec z^{\,2}-d^i_s\vec x\cdot\vec y+e^i_s\vec x\cdot\vec z+f^i_s\vec y\cdot\vec z\right) \, ,
\end{eqnarray}
\begin{eqnarray}
\label{wave2-4}
(AA)\Rightarrow R_4^i&=&
{\rm Exp}\left(-a^i_s\vec x^{\,2}-b^i_s\vec y^{\,2}-c^i_s\vec z^{\,2}-d^i_s\vec x\cdot\vec y-e^i_s\vec x\cdot\vec z-f^i_s\vec y\cdot\vec z\right)\\\nonumber
&-&{\rm Exp}\left(-a^i_s\vec x^{\,2}-b^i_s\vec y^{\,2}-c^i_s\vec z^{\,2}+d^i_s\vec x\cdot\vec y-e^i_s\vec x\cdot\vec z+f^i_s\vec y\cdot\vec z\right)\\\nonumber
&-&{\rm Exp}\left(-a^i_s\vec x^{\,2}-b^i_s\vec y^{\,2}-c^i_s\vec z^{\,2}+d^i_s\vec x\cdot\vec y+e^i_s\vec x\cdot\vec z-f^i_s\vec y\cdot\vec z\right)\\\nonumber
&+&{\rm Exp}\left(-a^i_s\vec x^{\,2}-b^i_s\vec y^{\,2}-c^i_s\vec z^{\,2}-d^i_s\vec x\cdot\vec y+e^i_s\vec x\cdot\vec z+f^i_s\vec y\cdot\vec z\right) \, .
\end{eqnarray}
The last equations can be expressed in a compact
manner by defining the following function,
\begin{equation}
\label{red1}
g(s_1,s_2,s_3)={\rm Exp}\left(-a^i_s\vec x^{\,2}-b^i_s\vec y^{\,2}-c^i_s\vec z^{\,2}
-s_1d^i_s\vec x\cdot\vec y-s_2e^i_s\vec x\cdot\vec z-s_3f^i_s\vec y\cdot\vec z\right),
\end{equation}
and the vectors
\begin{equation}
\vec{G}_s^i=\left(\begin{array}{l} g(+,+,+)\\g(-,+,-)\\g(-,-,+)\\g(+,-,-)\end{array}\right)\, ,
\end{equation}
and
\begin{eqnarray}
\label{red2}
\vec{\alpha}_{SS}&=&(+,+,+,+)\\ \nonumber
\vec{\alpha}_{SA}&=&(+,-,+,-)\\ \nonumber
\vec{\alpha}_{AS}&=&(+,+,-,-)\\ \nonumber
\vec{\alpha}_{AA}&=&(+,-,-,+),
\end{eqnarray}
which allows to write Eqs.~(\ref{wave2-2})--(\ref{wave2-4}) as,
\begin{eqnarray}
\label{redu}
(SS)&\Rightarrow& R_1^i=\vec{\alpha}_{SS}\cdot\vec{G}_s^i\\ \nonumber
(SA)&\Rightarrow& R_2^i=\vec{\alpha}_{SA}\cdot\vec{G}_s^i\\ \nonumber
(AS)&\Rightarrow& R_3^i=\vec{\alpha}_{AS}\cdot\vec{G}_s^i\\ \nonumber
(AA)&\Rightarrow& R_4^i=\vec{\alpha}_{AA}\cdot\vec{G}_s^i \, .
\end{eqnarray}
The radial wave function includes all possible internal relative orbital angular momenta
coupled to $L=0$. It has also well-defined symmetry properties on the $\vec z$ 
coordinate. Being $P_{(12)(34)}(\vec z \rightarrow -\vec z)R^i_{s_4}=P_zR^i_{s_4}$
one obtains,
\begin{eqnarray}
 \label{parz}
 P_{(12)(34)}R_1^i&=&+R_1^i\\ \nonumber
 P_{(12)(34)}R_2^i&=&-R_2^i\\ \nonumber
 P_{(12)(34)}R_3^i&=&-R_3^i\\ \nonumber
 P_{(12)(34)}R_4^i&=&+R_4^i \, .
\end{eqnarray}
To evaluate radial matrix elements we use the notation introduced in Eq.~(\ref{redu}): 
\begin{equation}
\label{ra1}
\br R_{\gamma}^i|f(x,y,z)|R_{\beta}^j\et=\int_V(\vec \alpha_{S_\gamma}\cdot \vec G^i_s)f(x,y,z)(\vec \alpha_{S_\beta}\cdot \vec G^j_{s'})dV=
\vec \alpha_{S_\gamma}\cdot F^{ij}\cdot\vec \alpha_{S_\beta}\, ,
\end{equation}
where $\gamma$ and $\beta$ stand for the symmetry of the radial 
wave function and $F^{ij}$ is a matrix whose element $(a,b)$ is
defined through, 
\begin{equation}
F^{ij}_{ab}=\int_V(\vec G_s^i)_a(\vec G^j_{s'})_bf(x,y,z)dV\, ,
\end{equation}
being $(\vec G_s^i)_a$ the component $a$ of the vector $\vec G_s^i$. 
From Eq.~(\ref{red1}) one obtains,
\begin{equation}
g(s_1,s_2,s_3)g(s'_1,s'_2,s'_3)={\rm Exp}\left(-a_{ij}\vec x^{\,2}-b_{ij}\vec y^{\,2}-c_{ij}\vec z^{\,2}
-\bar s_{ij}\vec x\cdot\vec y-\bar e_{ij}\vec x\cdot\vec z-\bar f_{ij}\vec y\cdot\vec z\right) \, ,
\end{equation}
where we have shortened the previous notation according to $a^i_s\to a_i$, 
$a_{ij}=a_i+a_j$ and $\bar d_{ij}=(s_1d_i+s_1'd_j)$. Therefore, 
all radial matrix elements will contain integrals of the form,
\begin{equation}
I=\int_V{\rm Exp}\left(-a_{ij}\vec x^{\,2}-b_{ij}\vec y^{\,2}-c_{ij}\vec
z^{\,2} -\bar s_{ij}\vec x\cdot\vec y-\bar e_{ij}\vec x\cdot\vec z-\bar f_{ij}\vec y\cdot\vec z\right)f(x,y,z)d\vec xd\vec yd\vec z  \, ,
\end{equation}
where the functions $f(x,y,z)$ are the potentials. Being
all of them radial functions (not depending on angular variables) 
one can solve the previous integral by noting:
\begin{equation}
\int {\rm Exp}\big[-\sum_{i,j=1}^nA_{ij}\vec x_i \cdot \vec x_j\big]f\big(|\sum \alpha_k\vec
x_k|\big)d\vec x_1...d\vec
x_n=\Bigg({\pi^n\over{det\,A}}\Bigg)^{3\over2}4\pi\Bigg({\Omega_{ij}\over\pi}\Bigg)^{3\over2}F(\Omega_{ij},f) \, ,
\end{equation}
where
\begin{eqnarray}
{1\over\Omega_{ij}}&=&\bar\alpha\cdot A^{-1} \cdot\alpha\\ \nonumber
F(A,f)&=&\int e^{-Au^2}f(u)u^2du\\ \nonumber
det\,A&>&0\\ \nonumber
{1\over\Omega_{ij}}&>&0 \, .
\end{eqnarray}
One can extract some useful relations for the radial matrix elements using
simple symmetry properties. Let us rewrite Eq.~(\ref{ra1})
\begin{eqnarray}
\br R_{\gamma}^i|f(x,y,z)|R_{\beta}^j\et&=&\br R_{P_xP_yP_z}^i|f(x,y,z)|R_{P_x'P_y'P_z'}^j\et\\ \nonumber
&=&\int_x\int_y\int_z R_{P_xP_yP_z}^if(x,y,z)R_{P_x'P_y'P_z'}^jd\vec xd\vec yd\vec z \, .
\end{eqnarray}
If $f(x,y,x)$ depends only in one coordinate, for example $\vec x$, the 
integrals over the other coordinates will be zero if one of them has different symmetry properties,
$P_y\neq P_y'$ or $P_z\neq P_z'$ in our example. Therefore
\begin{eqnarray}
\br R_{\gamma}^i|f(x)|R_{\beta}^j\et&\propto&\delta_{\gamma\beta}\\ \nonumber
\br R_{\gamma}^i|f(y)|R_{\beta}^j\et&\propto&\delta_{\gamma\beta}\\ \nonumber
\br R_{\gamma}^i|f(z)|R_{\beta}^j\et&\propto&\delta_{\gamma\beta}\\ \nonumber
\br R_{\gamma}^i|{\rm Constant}|R_{\beta}^j\et&\propto&\delta_{\gamma\beta} \, .
\end{eqnarray}
The radial wave function described in this section is adequate to describe 
not only bound states, but also it is flexible enough to describe states of 
the continuum within a reasonable accuracy~\cite{Suz98,Vij09,Via09}.

\section{Two--body amplitudes}
\label{secIV}

We have constructed the two-body amplitudes for all subsystems entering the 
three- and four-body problems studied by solving the Lippmann--Schwinger
equation of each $(i,j)$ channel,
\begin{equation}
t^{ij}(p,p';e)= V^{ij}(p,p')+\int_0^\infty {p^{\prime\prime}}^2
dp^{\prime\prime} V^{ij}(p,p^{\prime\prime})
\frac{1}{e-{p^{\prime\prime}}^2/2\mu} t^{ij}(p^{\prime\prime},p';e) \, ,
\label{eq19} 
\end{equation}
where 
\begin{equation}
V^{ij}(p,p')=\frac{2}{\pi}\int_0^\infty r^2dr\; j_0(pr)V^{ij}(r)j_0(p'r) \, ,
\label{eq20} 
\end{equation}
and the two-body potentials consist of an attractive and a repulsive
Yukawa term, i.e.,
\begin{equation}
V^{ij}(r)=-A\frac{e^{-\mu_Ar}}{r}+B\frac{e^{-\mu_Br}}{r} \, .
\label{eq21} 
\end{equation}
The parameters of the $\Lambda N$, $\Xi N$, $\Lambda\Lambda$ and $\Xi \Xi$ 
channels were obtained by fitting the low-energy data and the phase shifts of 
each channel as given by the most recent update of the strangeness $-1$~\cite{Nae15}, $-2$~\cite{Nag15} 
and $-3$ and $-4$~\cite{Rij13} ESC08c Nijmegen 
potentials. In the case of the $NN$ interaction we use
the Malfliet-Tjon models~\cite{Mal69} with the parameters given in Ref.~\cite{Gib90}.
The low-energy data and the parameters of these models are given in Table~\ref{t1}. 
It is worth to note that the scattering length and effective range of the most recent
update of the $\Lambda\Lambda$ interaction derived from chiral effective field
theories are very much like those of the ESC08c Nijmegen potential (see Table 2 of
Ref.~\cite{Hai16}) unlike the earlier version 
used in Ref.~\cite{Ric15} (see Table 4 of Ref.~\cite{Pol07})
reporting remarkably small effective ranges.

The $\Xi N$ $^1S_0$ $(I=0)$ potential was fitted to the $\Xi N$ phase shifts given in Fig. 14 
of Ref.~\cite{Nag15} without taking into account the inelasticity, i.e., assuming $\rho=0$
(this two-body channel does not contribute to the three- and four-body bound states found
in this work). Regarding the two-body interactions containing a 
single $\Lambda$, they are constrained by a simultaneous fit
to the combined $NN$ and $YN$ scattering data, supplied with 
constraints on the $YN$ and $YY$ interaction originating from 
the G-matrix information on hypernuclei~\cite{Nae15}.

\begin{table}[t]
\caption{Low-energy data and parameters of the local central Yukawa-type potentials 
given by Eq.~(\ref{eq21}) for the $NN$ potential~\cite{Gib90},
and the most recent updates of the ESC08c Nijmegen interactions
for the $\Lambda N$~\cite{Nae15}, $\Xi N$~\cite{Nag15},
$\Xi \Xi$~\cite{Rij13}, and $\Lambda \Lambda$~\cite{Nag15} systems.} 
\begin{ruledtabular} 
\begin{tabular}{ccccccccc} 
& $(i,j)$ & $a({\rm fm})$ & $r_0({\rm fm})$ & $A$(MeV fm) & 
$\mu_A({\rm fm}^{-1}$) 
& $B$(MeV fm) & $\mu_B({\rm fm}^{-1})$  & \\
\hline
\multirow{1}{*}{$NN$} & $(1,0)$  & $-23.56$ & $2.88$ & $513.968$  & $1.55$  & $1438.72$ & $3.11$ & \\
\hline
\multirow{2}{*}{$\Lambda N$}& $(1/2,0)$ & $-2.62$  & $3.17$  &  $416$  & $1.77$  & $1098$ & $3.33$ & \\ 
& $(1/2,1)$ & $-1.72$  & $3.50$  &  $339$  & $1.87$  & $968$ & $3.73$ & \\ \hline
\multirow{4}{*}{$\Xi N$}
& $(0,0)$\footnotemark[1] & $-$      & $-$      &  $120$  & $1.30$  & $510$ & $2.30$ & \\
& $(0,1)$ & $-5.357$ & $1.434$  &  $377$  & $2.68$  & $980$ & $6.61$ & \\ 
& $(1,0)$ & $0.579$  & $-2.521$ &  $290$  & $3.05$  & $155$ & $1.60$ & \\
& $(1,1)$ & $4.911$  & $0.527$  &  $568$  & $4.56$  & $425$ & $6.73$ & \\ \hline
\multirow{2}{*}{$\Xi \Xi$} & $(0,1)$ & $0.53$   & $1.63$  &  $210$  & $1.60$  & $560$ & $2.05$ & \\ 
& $(1,0)$ & $-7.25$  & $2.00$  &  $155$  & $1.75$  & $490$ & $5.60$ & \\ \hline
\multirow{1}{*}{$\Lambda \Lambda$} & $(0,0)$ & $-0.853$& $5.126$ &  $121$  & $1.74$  & $926$ & $6.04$ & \\ 
\end{tabular}
\footnotetext[1]{This channel is discussed on Sec.~\ref{secIV}.}
\end{ruledtabular}
\label{t1} 
\end{table}
\begin{figure*}[t]
\resizebox{8.cm}{12.cm}{\includegraphics{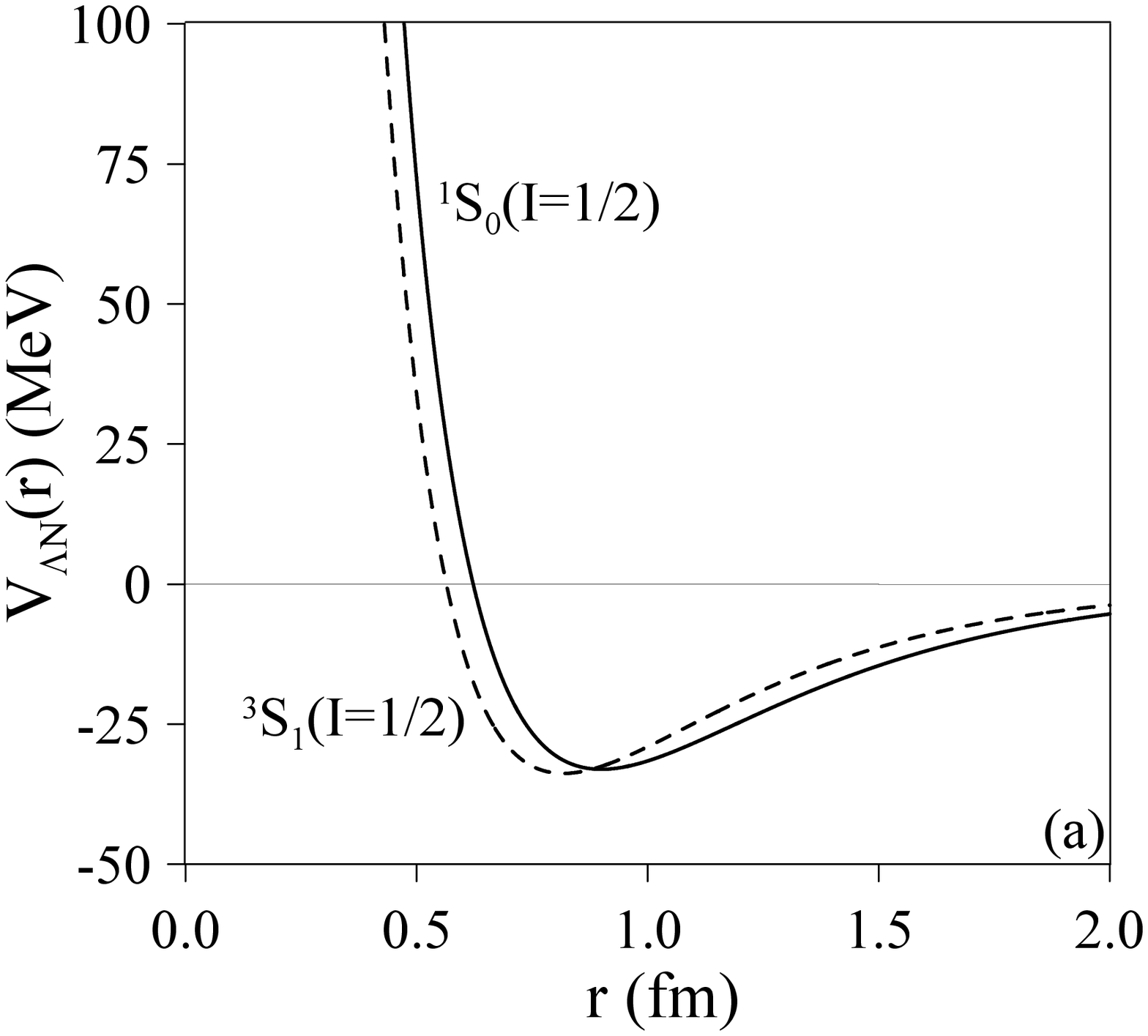}}
\resizebox{8.cm}{12.cm}{\includegraphics{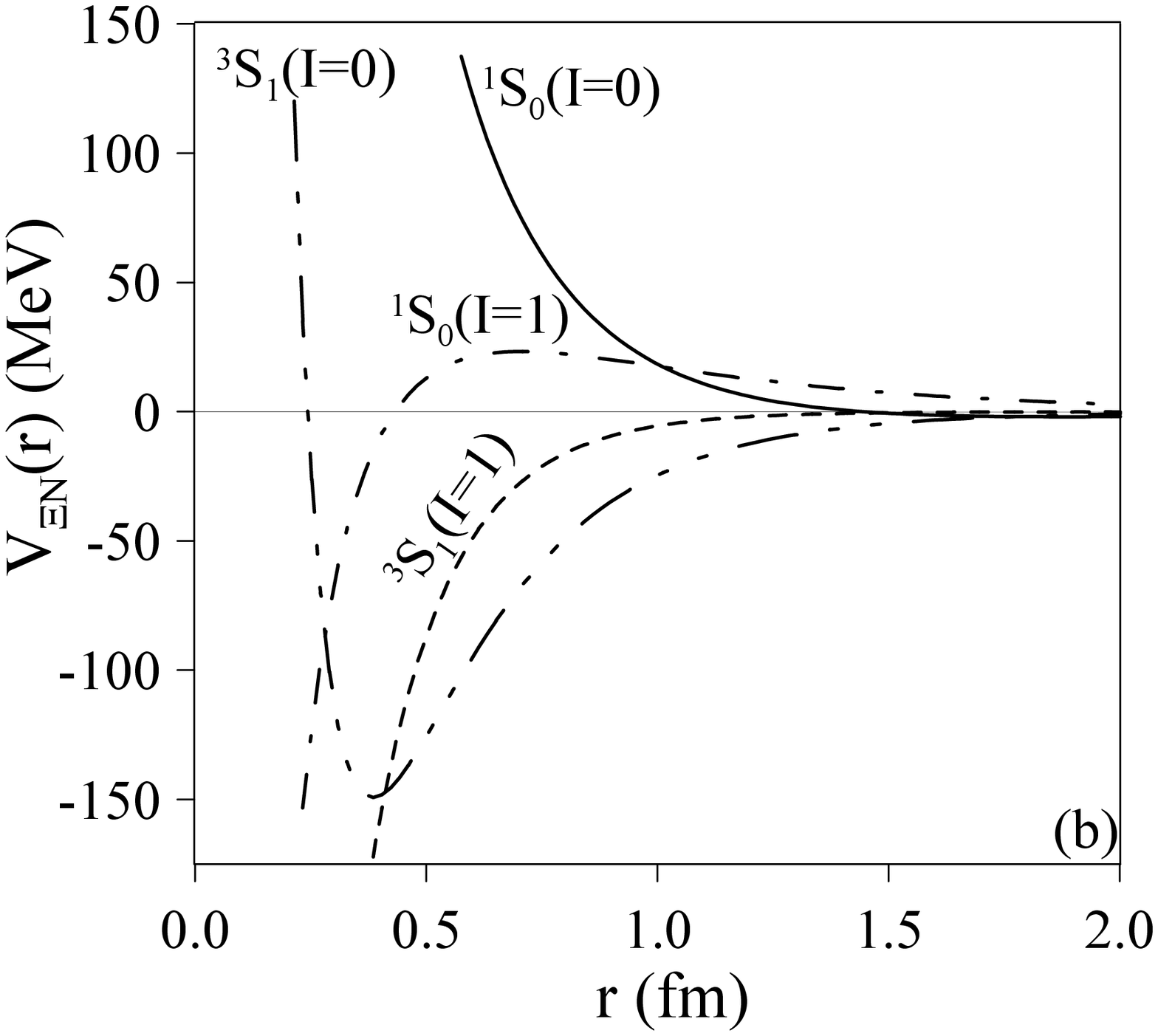}}\vspace*{-5.0cm}
\resizebox{8.cm}{12.cm}{\includegraphics{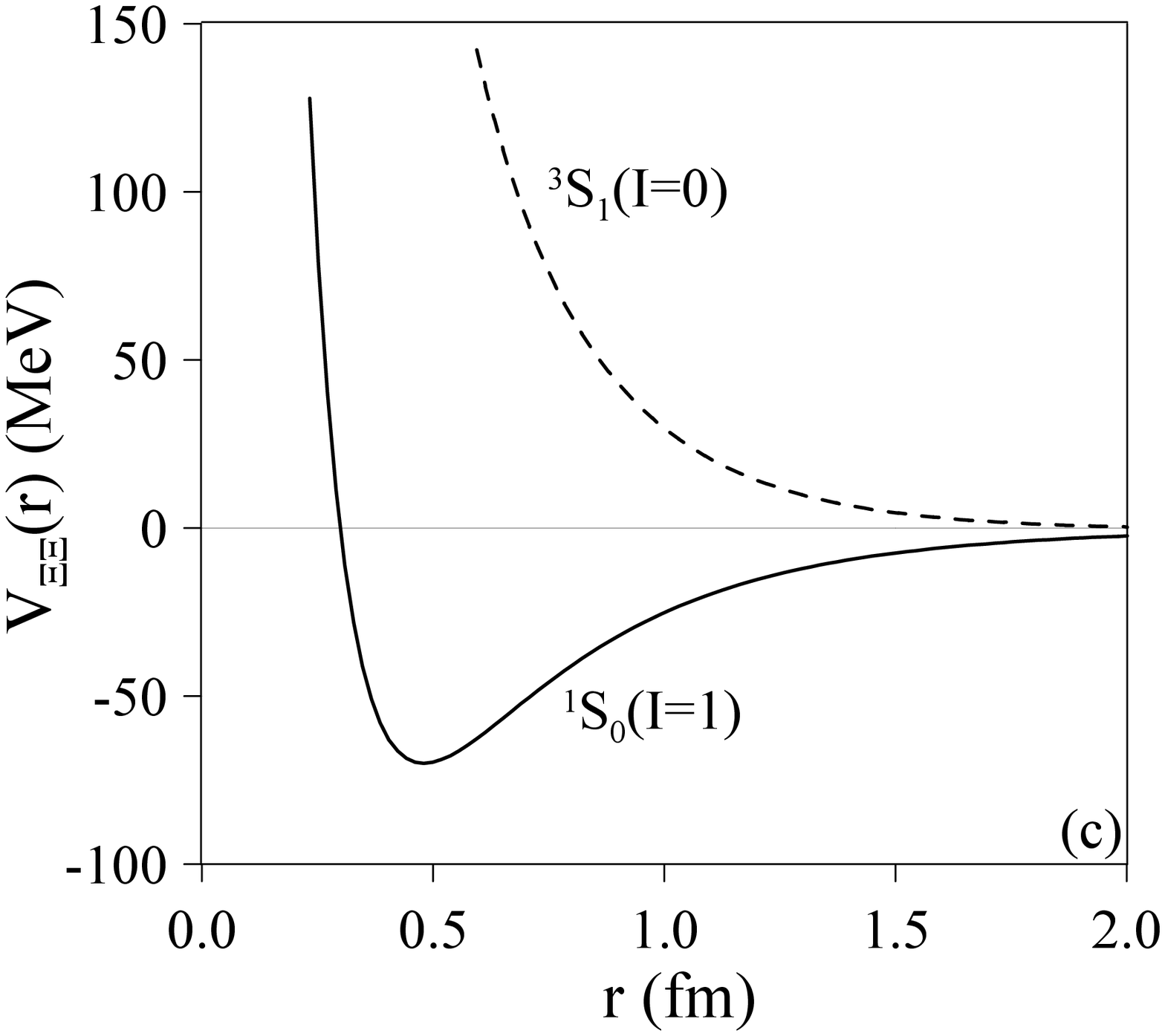}}
\resizebox{8.cm}{12.cm}{\includegraphics{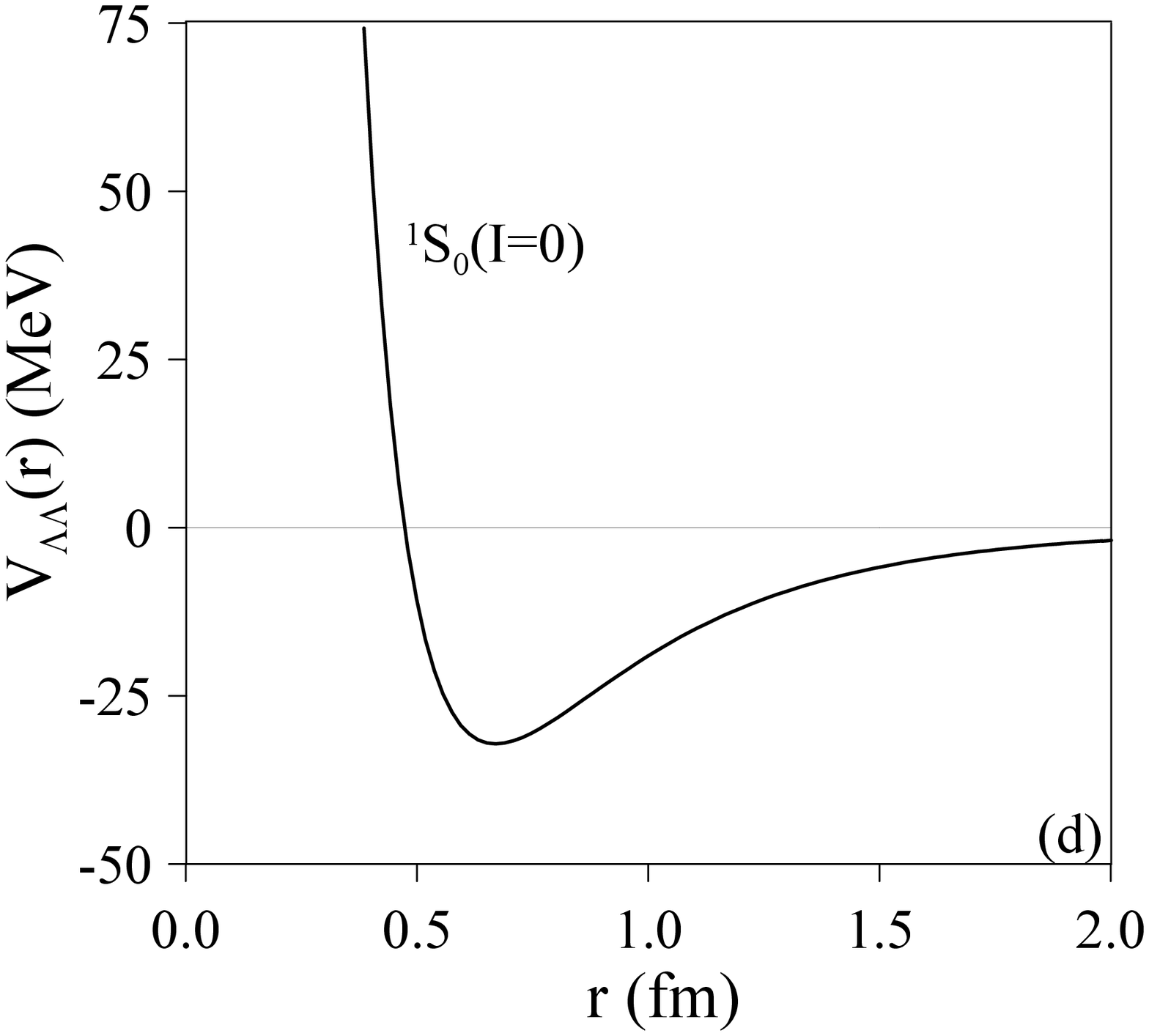}}
\vspace*{-5.0cm}
\caption{(a) $V_{\Lambda N}(r)$ potential as given by Eq.~(\ref{eq21}) with the
parameters of Table~\ref{t1}.
(b) Same as (a) for the $V_{\Xi N}(r)$ potential.
(c) Same as (a) for the$V_{\Xi\Xi}(r)$ potential.
(d) Same as (a) for the $V_{\Lambda\Lambda}(r)$ potential.}
\label{fig1}
\end{figure*}

The potentials obtained are shown in Fig.~\ref{fig1}. In Fig.~\ref{fig1}(a) we show the
$V_{\Lambda N}(r)$ potential that it is tightly constrained by the existing experimental data.
The interaction is attractive at intermediate range and strongly repulsive at short range,
but without having bound states. 
In Fig.~\ref{fig1}(b) we show the $V_{\Xi N}(r)$ potential,
where one notes the attractive character of the $^3S_1(I=1)$ $\Xi N$ partial wave, 
giving rise to the $D^*$ bound state~\cite{Naa15} with a binding energy of 1.6 MeV. 
We also confirm how all the $J=1$ and $I=1$ $\Xi N$ interactions are attractive~\footnote{
There are also models for the strangeness $-2$ baryon-baryon interaction
based on EFT calculations~\cite{Pol07} showing $I=1$ $\Xi N$ attraction, although 
one cannot conclude the strength of the
interaction due to the huge effective ranges reported.}~\cite{Rij13}.
Regarding the $\Xi\Xi$ interaction, Fig.~\ref{fig1}(c), we
observe the attractive character of the $^1S_0(I=1)$ potential, that although having 
bound states in earlier versions of the ESC08c Nijmegen potential~\cite{Sto99}, in the most recent
update of the strangeness $-4$ sector it does not present a bound state~\cite{Rij13}.
The existence of bound states in the $\Xi\Xi$ system has been predicted by different 
calculations in the literature~\cite{Bea12,Mil06,Hai10}. It can be definitively stated 
that all models agree on the fairly important attractive character of this 
channel, either with or without a bound state~\cite{Hai15}.
Finally, in Fig.~\ref{fig1}(d) we show the $V_{\Lambda\Lambda}(r)$ potential,
mainly determined by the $NN$ and $YN$ data, and SU(3)
symmetry~\cite{Nag15,Rij16}. It gives account of the pivotal
results of strangeness $-2$ physics, the NAGARA~\cite{Tak01} 
and the KISO~\cite{Naa15} events. Although other double-$\Lambda$
hypernuclei events, like the DEMACHIYANAGI and HIDA events~\cite{Nak10},
are not explicitly taken into account, the G-matrix nuclear matter study
of $\Xi^-$ capture both in $^{12}$C and $^{14}$N (see section
VII of Ref.~\cite{Nag15}), concludes that the $\Xi N$ attraction
in the ESC08c potential is consistent with the $\Xi$-nucleus binding 
energies given by the emulsion data of the twin $\Lambda$-hypernuclei.

\section{Results and discussion}
\label{secV}
\begin{figure*}[t]
\resizebox{8.cm}{12.cm}{\includegraphics{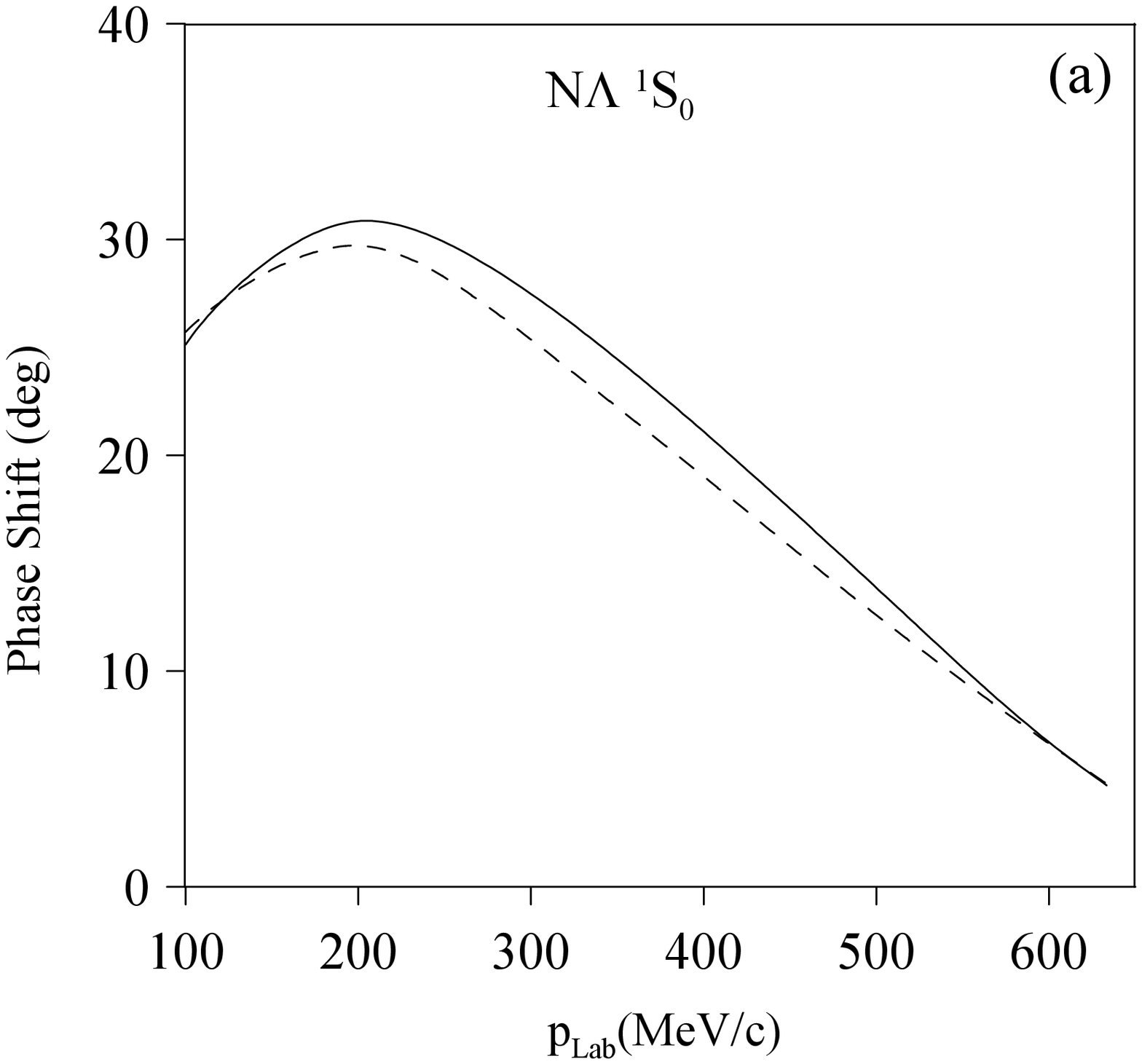}}
\resizebox{8.cm}{12.cm}{\includegraphics{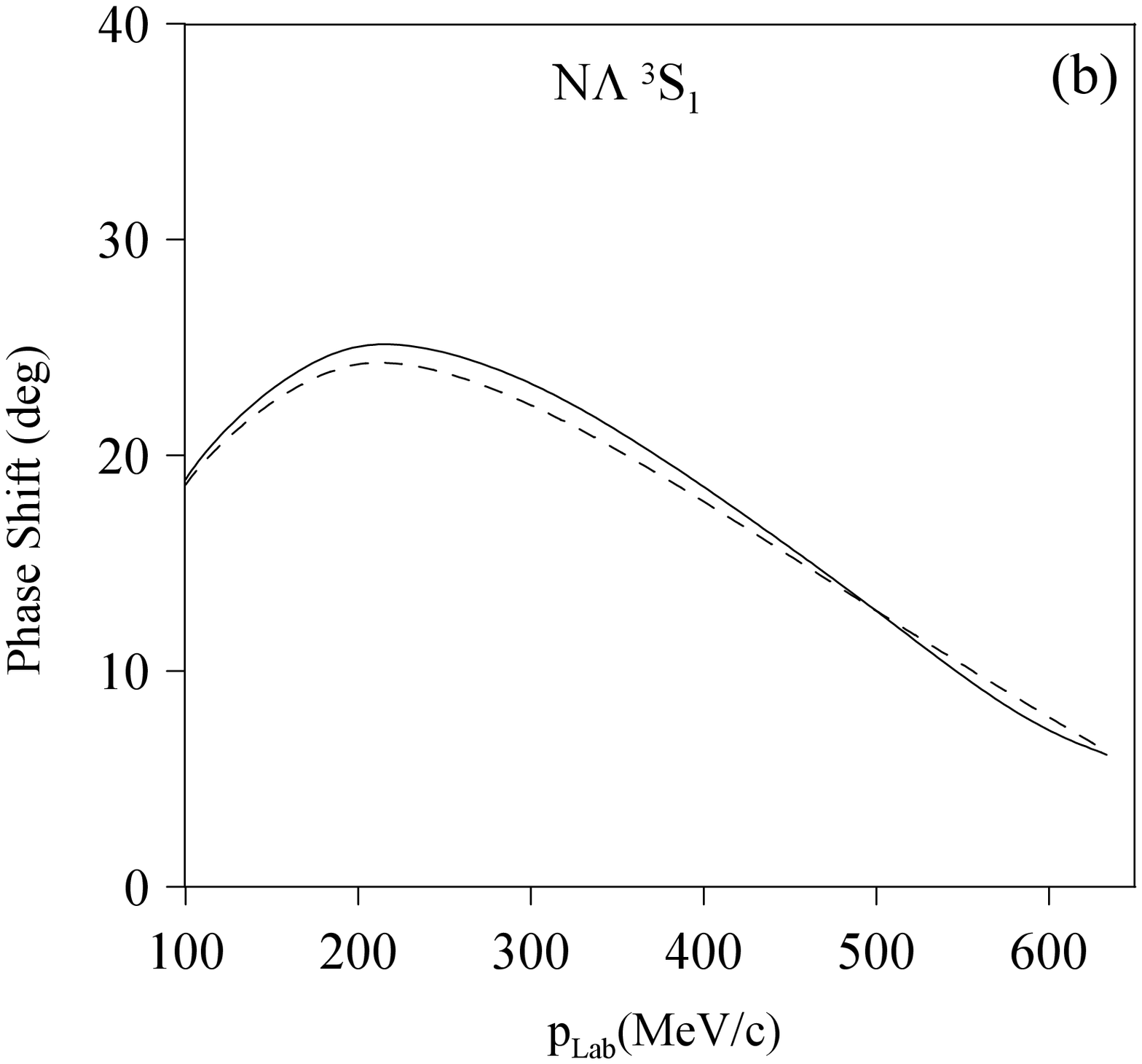}}\vspace*{-5.0cm}
\resizebox{8.cm}{12.cm}{\includegraphics{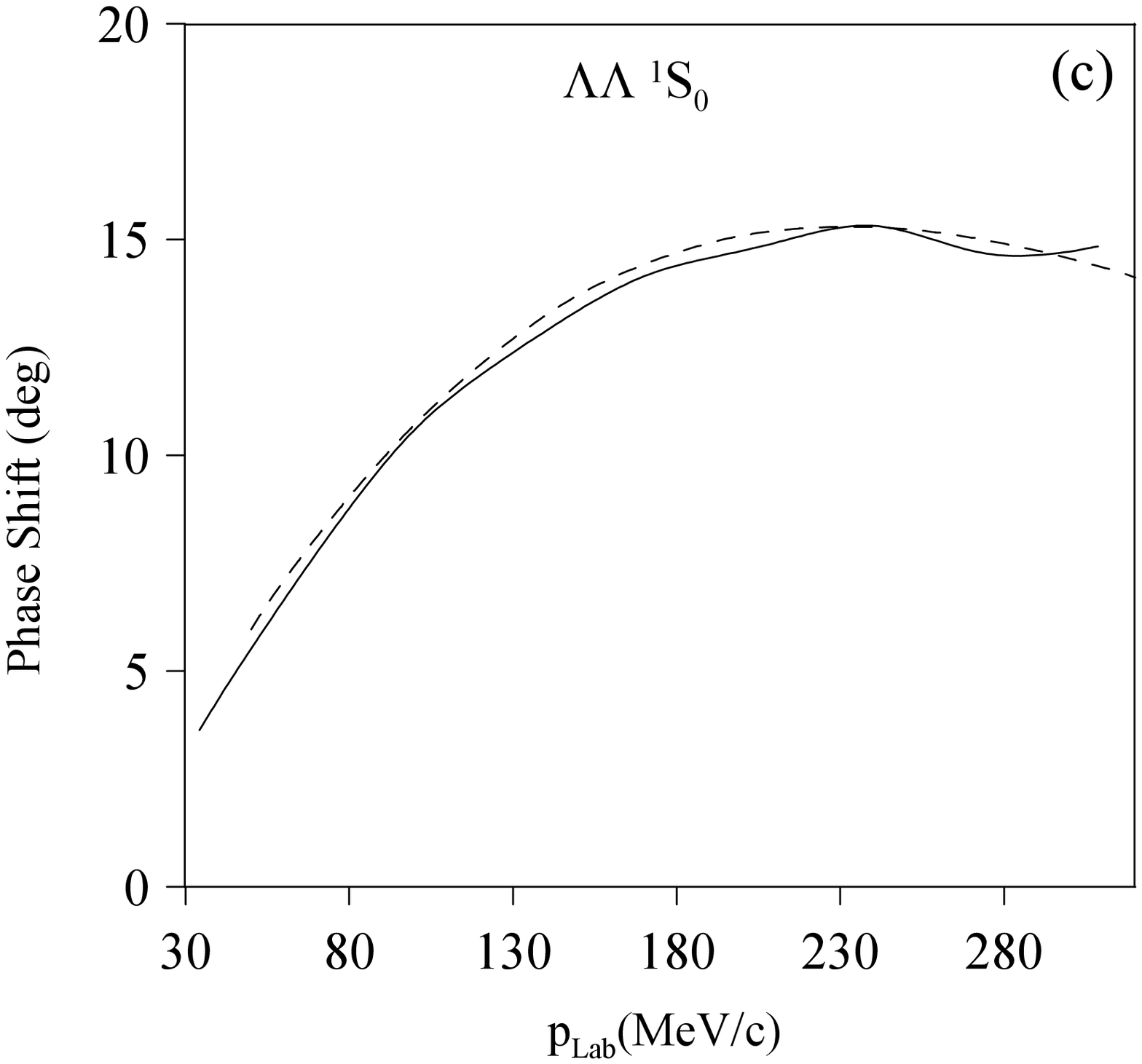}}
\vspace*{-5.0cm}
\caption{(a) $\Lambda N$ $^1S_0$ phase shifts. The solid line stands for the
results of the ESC08c Nijmegen potential and the dashed line for the results
of the two-body potential of Eq.~(\ref{eq21}) with the parameters given in 
Table~\ref{t1}. (b) Same as (a) for the $\Lambda N$ $^3S_1$ phase shifts.
(c) Same as (a) for the $\Lambda\Lambda$ $^1S_0$ phase shifts.}
\label{fig2}
\end{figure*}

Let us first of all show the reliability of the input potentials.
We compare in Fig.~\ref{fig2} the $\Lambda N$ and $\Lambda\Lambda$
phase shifts reported by the ESC08c Nijmegen
potential and those obtained by our fits with the two-body potentials of 
Eq.~(\ref{eq21}) and the parameters given in Table~\ref{t1}. As can be
seen the agreement is good. As stated above, the $\Xi N$ $^1S_0$ $(I=0)$ 
potential was fitted to the $\Xi N$ phase shifts given in Fig. 14 
of Ref.~\cite{Nag15}. Once we have described the phase shifts,
the $\Lambda N$ and $\Lambda\Lambda$ potentials include in an effective 
manner the coupling to other two-body channels as it may be the $\Sigma N$
or $\Xi N$ two-body systems\footnote{Although by fitting the $\Lambda N$ 
phase shifts, the coupling to the 
$\Sigma N$ system has been included in an effective manner,
it would also be interesting to unfold the effective $\Lambda N$ interaction,
separating the contribution from $\Lambda N \leftrightarrow \Sigma N$.
As it has been discussed in the literature ~\cite{Gar14,Hiy14,Gar07,Miy95,Gib77,Gib79} 
the hypertriton does not get bound by considering only $\Lambda NN$ channels, but it is 
necessary to include also $\Sigma NN$ channels.
Similar considerations hold for the $\Lambda\Lambda \leftrightarrow \Xi N$ coupling.}.
We have also tested the two-body interactions in the three-body problem of 
systems made of $N$'s and $\Lambda$'s.
The hypertriton is bound by 144 keV, and the $\Lambda nn$ system is unbound.

The reasonable description of the known two- and three-body problems gives confidence
to address the study of other three- and four-body systems.
We show in Table~\ref{t2} the channels of the different two-body subsystems contributing
to each $(I,J)$ three- and four-body state that we will study. For the $\Xi\Xi NN$ system we only consider
the $I=2$ channels, because the $I=0$ and $1$ states would decay strongly to $\Lambda\Lambda NN$ states. 
The three- and four-body problems are studied by means
of the ESC08c Nijmegen interactions described in Sec.~\ref{secIV} and given in Table~\ref{t1}. 
The binding energies are measured with respect to the lowest threshold, indicated in Table~\ref{t2} 
for each particular state. 
\begin{table}[t]
\caption{Two-body $NN$, $YN$ and $YY$ isospin-spin $(i,j)$ channels that contribute to a 
given three- or four-body state with total isospin $I$ and total spin $J$. The last column
indicates the corresponding threshold for each state, that would come given by $\sum_{i=1}^{3(4)} M_i - E$,
where $M_i$ are the masses of the baryons of each channel, $B_1$ stands for the binding energy of 
the deuteron and $B_2$ for the binding energy of the $D^*$ $\Xi N$ state.}
\begin{ruledtabular} 
\begin{tabular}{cccccccc} 
& $(I,J)$ & $\Lambda N$ & $\Xi N$ &  $\Xi \Xi(NN)$ & $\Lambda\Lambda$ & $E$ \\
\hline
\multirow{4}{*}{$\Xi NN$}
& $(1/2,1/2)$ & $-$ & (0,0),(0,1),(1,0),(1,1) & (0,1),(1,0) & $-$ & $B_1$ \\
& $(1/2,3/2)$ & $-$ & (0,1),(1,1) & (0,1) & $-$ &  $ B_1$ \\
& $(3/2,1/2)$ & $-$ & (1,0),(1,1) & (1,0) & $-$ &$ B_2$ \\
& $(3/2,3/2)$ & $-$ & (1,1) & $-$ &  $-$ & $ B_2$ \\ \hline
\multirow{4}{*}{$\Xi \Xi N$} 
& $(1/2,1/2)$ & $-$ & (0,0),(0,1),(1,0),(1,1) & (0,1),(1,0) & $-$ &$ B_2 $ \\
& $(1/2,3/2)$ & $-$ & (0,1),(1,1) & (0,1) &  $-$ & $ B_2$  \\
& $(3/2,1/2)$ & $-$ & (1,0),(1,1) & (1,0) & $-$ & $ B_2$ \\
& $(3/2,3/2)$ & $-$ & (1,1) & $-$ & $-$ & $ B_2$ \\ \hline
\multirow{1}{*}{$\Xi \Xi NN$} 
& $(2,0)$ & $-$ & (1,0),(1,1) &  (1,0) & $-$ & $ 2 B_2 $ \\ \hline
\multirow{1}{*}{$\Lambda \Lambda NN$} 
& $(1,0)$ & (1/2,0),(1/2,1) & $-$ &  (1,0) & (0,0) & $ 0 $ \\
\end{tabular}
\end{ruledtabular}
\label{t2} 
\end{table}
\subsection{Three-body systems}
\label{secVa}

\begin{figure*}[t]
\resizebox{8.cm}{12.cm}{\includegraphics{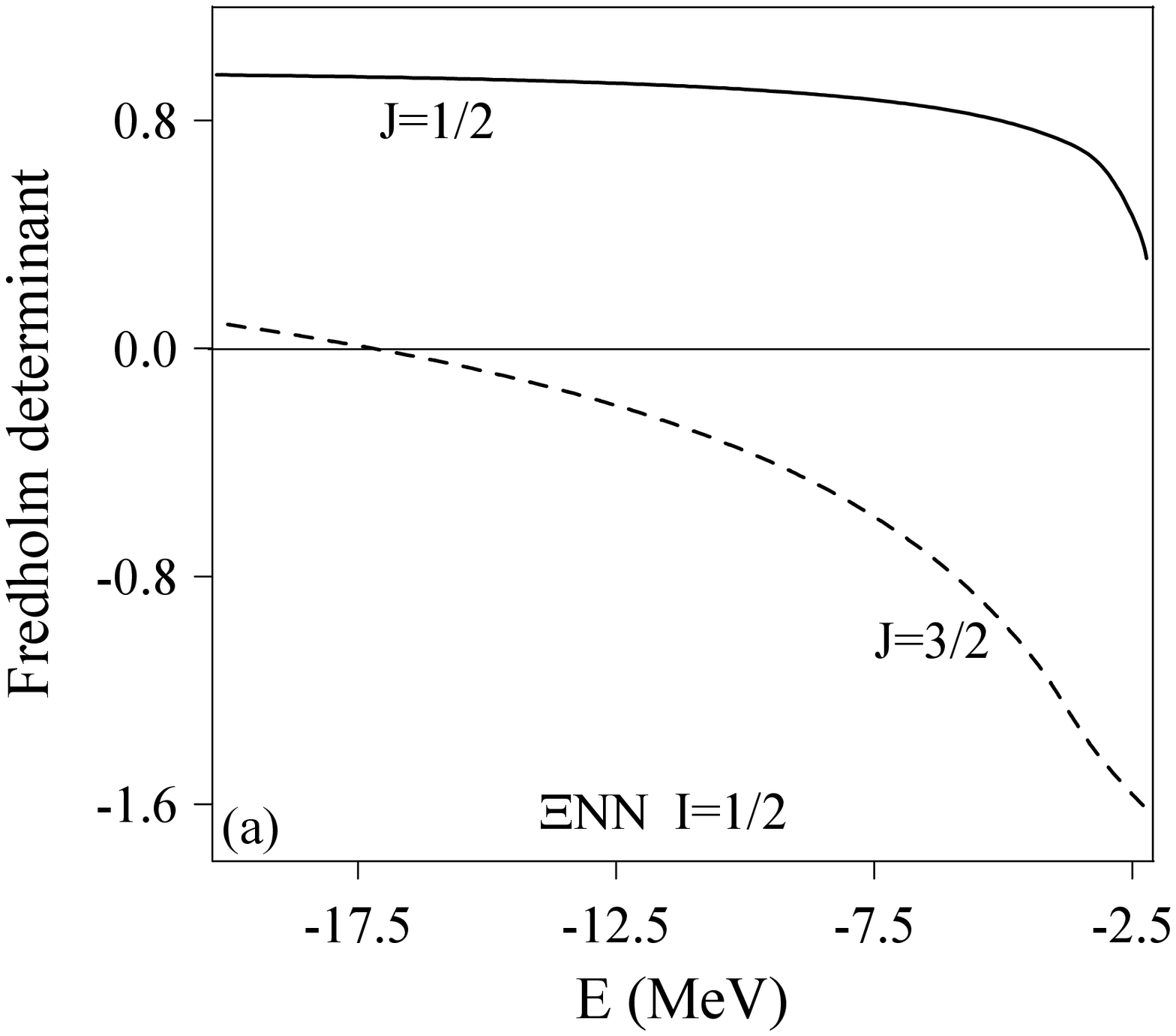}}
\resizebox{8.cm}{12.cm}{\includegraphics{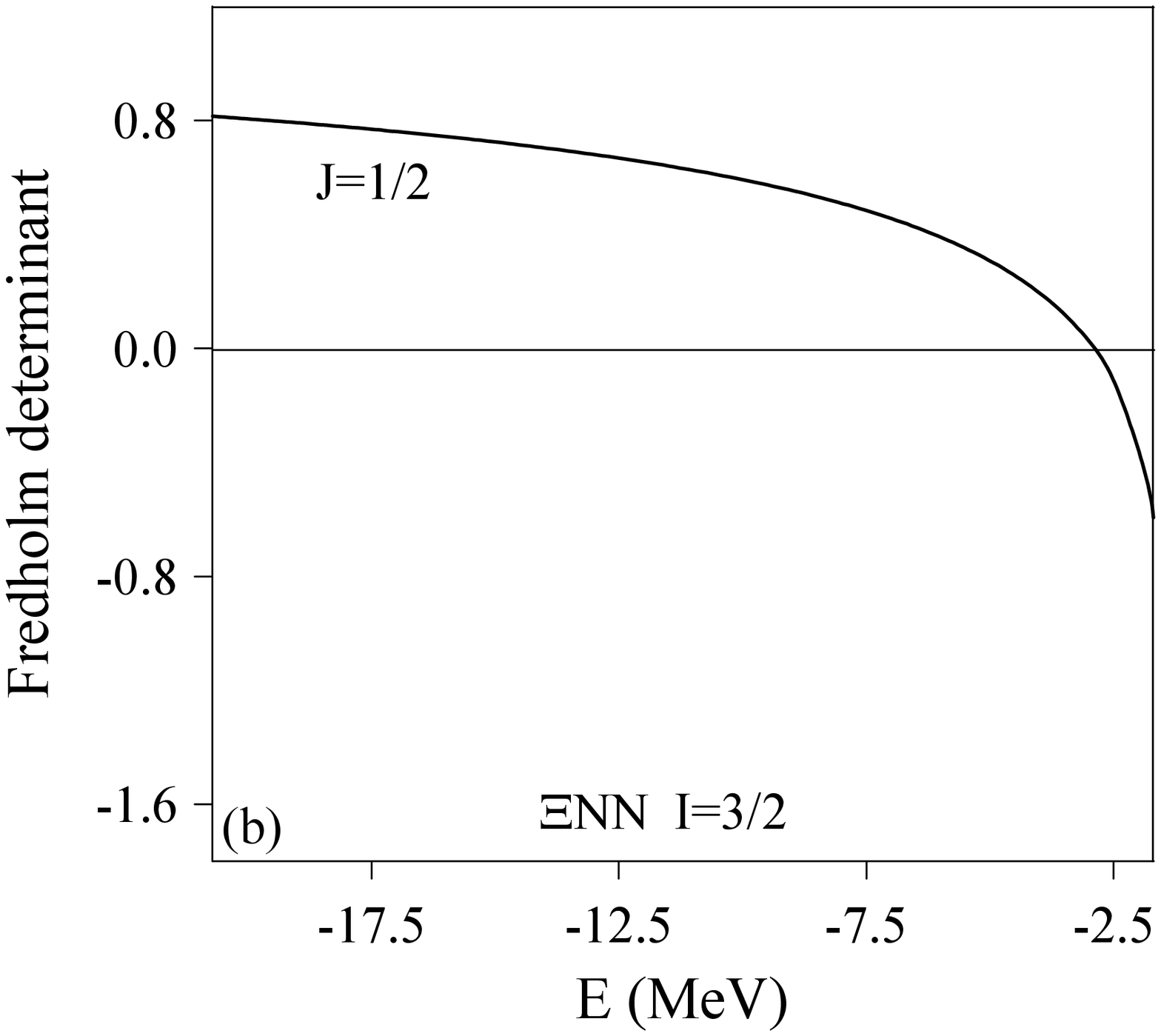}}
\vspace*{-5.0cm}
\caption{(a) Fredholm determinant for the $J=1/2$ and $J=3/2$ $I=1/2$
$\Xi NN$ channels.
(b) Fredholm determinant for the $J=1/2$ $I=3/2$
$\Xi NN$ channel.}
\label{fig3}
\end{figure*}

We show in Fig.~\ref{fig3} the Fredholm determinant
of all $\Xi NN$ channels~\cite{Gar15,Gar16}.
As we can see in Fig.~\ref{fig3}(b), a bound state is found for the
$(I)J^P=(\frac{3}{2})\frac{1}{2}^+$ $\Xi NN$ state, 1.3 MeV below
the corresponding threshold, $2m_N+m_\Xi-B_2$, where $B_2$ is the binding energy 
of the $D^*$ $\Xi N$ state. However, the most interesting result 
of the $\Xi NN$ system is shown in Fig.~\ref{fig3}(a), the very large binding
energy of the $(\frac{1}{2})\frac{3}{2}^+$ state, which would make it easy to identify 
experimentally as a sharp resonance lying some $17.2$ MeV below the $\Xi NN$ threshold.
The $\Lambda\Lambda - \Xi N$ $(i,j)=(0,0)$ transition channel, which is
responsible for the decay $\Xi NN\to\Lambda\Lambda N$, does not contribute to the 
$(I)J^P=(\frac{1}{2})\frac{3}{2}^+$ state in a pure $S-$wave configuration~\cite{Gar16}.
One would need at least the spectator nucleon to be in a $D-$wave or that the
$\Lambda\Lambda - \Xi N$ transition channel be in one of the 
negative parity $P-$wave channels, with the nucleon spectator also in a $P-$wave. 
Thus, due to the angular momentum barriers the resulting
decay width of the $(\frac{1}{2})\frac{3}{2}^+$ state is expected to be very small. 

\begin{figure*}[b]
\resizebox{8.cm}{12.cm}{\includegraphics{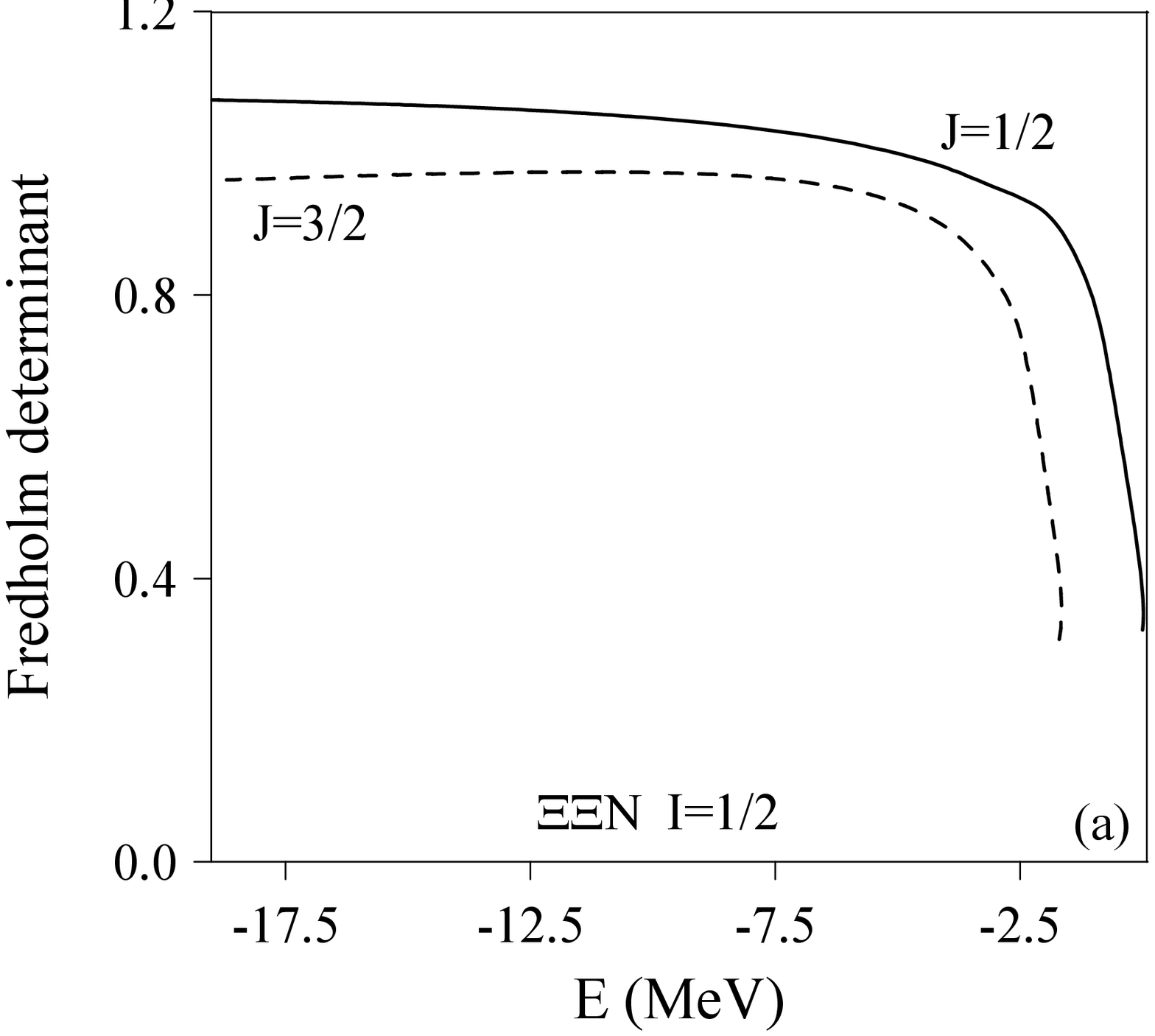}}
\resizebox{8.cm}{12.cm}{\includegraphics{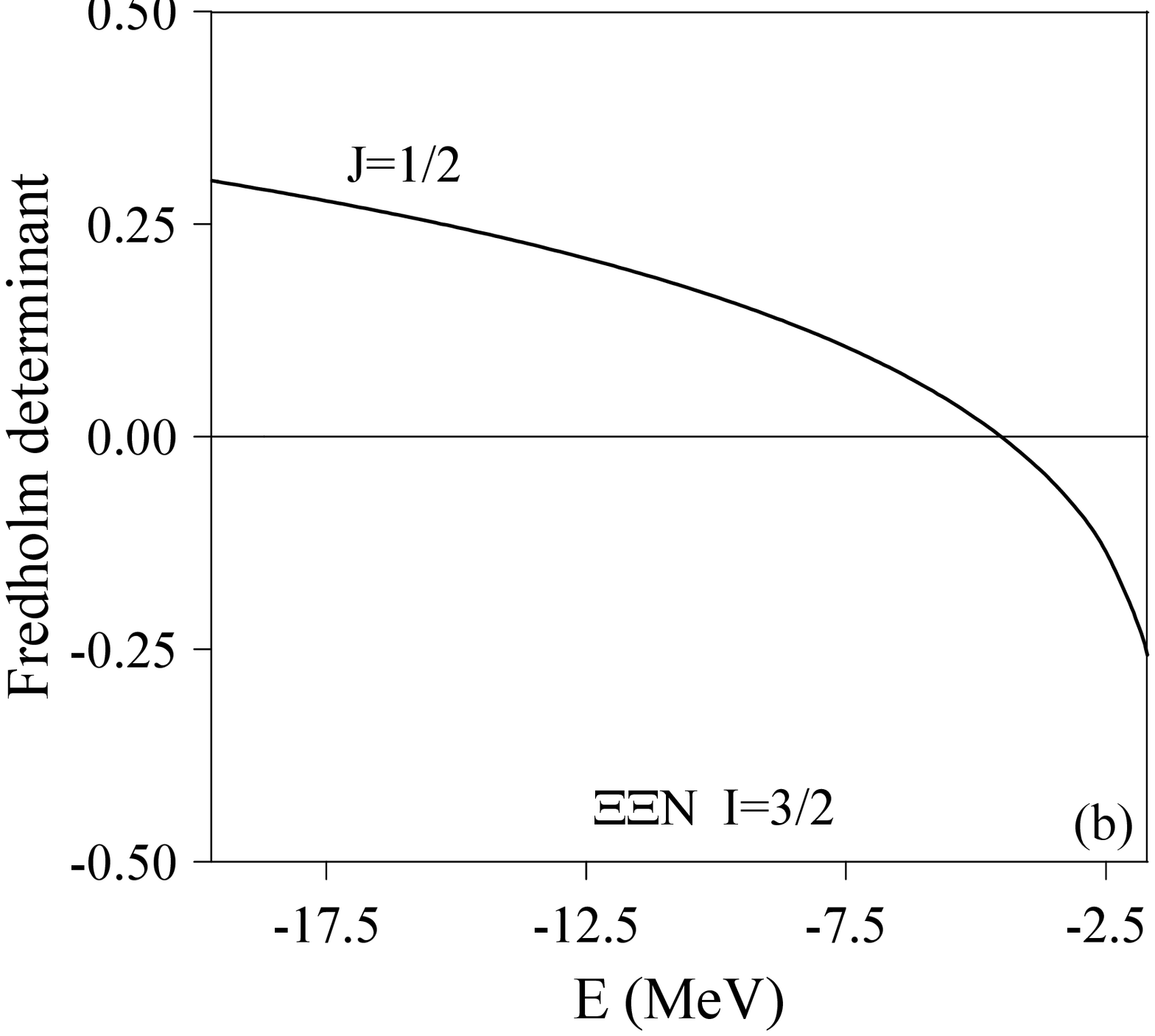}}
\vspace*{-5.0cm}
\caption{(a) Fredholm determinant for the $J=1/2$ and $J=3/2$ $I=1/2$
$\Xi\Xi N$ channels.
(b) Fredholm determinant for the $J=1/2$ $I=3/2$
$\Xi\Xi N$ channel.}
\label{fig4}
\end{figure*}

For the $\Xi NN$ three-baryon system with $(I,J)=(3/2,3/2)$, only the $(i,j)=(1,1)$ $\Xi N$ 
channel contributes (see Table~\ref{t2}),
and the corresponding Faddeev equations with two identical fermions 
can be written as~\cite{Gar07},
\begin{equation}
T=- \,t_N^{N\Xi} \, G_0 \, T \, .
\label{eqR}
\end{equation}
Thus, due to the negative sign in the r.h.s. the $\Xi N$ interaction is 
effectively repulsive and, therefore, no bound state is possible in spite of 
the attraction of the $\Xi N$ subsystem. The minus sign in 
Eq.~(\ref{eqR}) is a consequence of the identity of the two nucleons since 
the first term of the r.h.s. of Eq.~(\ref{eqR})
proceeds through $\Xi$ exchange and it corresponds to a diagram where the initial and final states 
differ only in that the two identical fermions have been interchanged which brings the minus sign. 
This effect has been pointed out before~\cite{Gar87}. This is the reason why the Fredholm 
determinant for the $(I,J)=(3/2,3/2)$ $\Xi NN$ channel is not shown in Fig.~\ref{fig3}(b).

Finally, we show in Fig.~\ref{fig4} the Fredholm determinant
of all $\Xi \Xi N$ channels. The Fredholm 
determinant for the $(I)J^P=(3/2)3/2^+$ channel is not shown in Fig.~\ref{fig4}(b)
for the same reason explained above for the $\Xi NN$ system, it is strongly repulsive.
In the $\Xi\Xi N$ system there appears a bound state with quantum numbers 
$(I)J^P=(\frac{3}{2})\frac{1}{2}^+$, 2.9 MeV below the lowest
threshold, $2m_\Xi + m_N - B_2$, where $B_2$ stands for the binding 
energy of the $D^*$ $\Xi N$ subsystem. 
Since this $\Xi\Xi N$ state has isospin $3/2$ it can not decay into
$\Xi\Lambda\Lambda$ due to isospin conservation so that it would be stable.
This stable state appears in spite of the fact
that the last update of the ESC08c Nijmegen $\Xi\Xi$ $^1S_0(I=1)$ potential 
has not bound states, as it is however predicted by several models
in the literature. If bound states would exist for the $\Xi \Xi$
system the three-body state would become deeply bound as it happens for the $\Xi NN$ system.
The $I=1/2$ channels are also attractive but they are not bound.

Let us finally mention that our results for three-body systems containing a $\Xi N$
subsystem has been recently reproduced by means of the configuration-space
Faddeev equations~\cite{Fil17}.
\subsection{Four-body systems}
\label{secVb}
\begin{figure*}[t]
\resizebox{8.cm}{12.cm}{\includegraphics{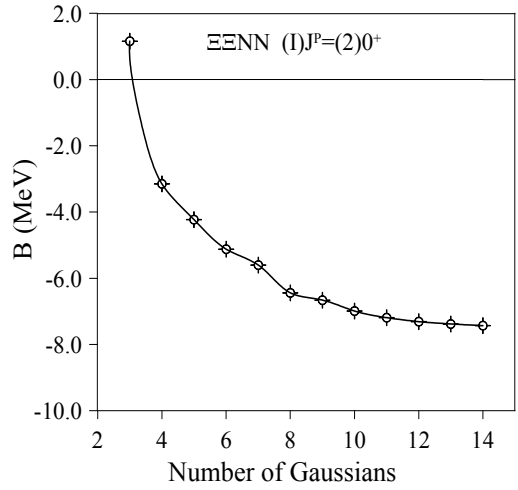}}
\vspace*{-5.0cm}
\caption{Binding energy of the $(I)J^P=(2)0^+$ $\Xi\Xi NN$ state as a function of
the number of Gaussians in the variational calculation.}
\label{fig5}
\end{figure*}
\begin{figure*}[b]
\resizebox{8.cm}{12.cm}{\includegraphics{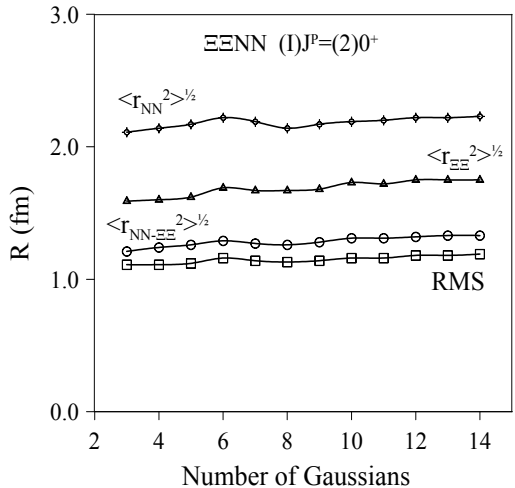}}
\vspace*{-5.0cm}
\caption{Root mean square radii of the $(I)J^P=(2)0^+$ $\Xi\Xi NN$ state as a function of
the number of Gaussians in the variational calculation. See text for details.}
\label{fig6}
\end{figure*}
In the previous section we have seen that all three-body systems made of $N$'s 
and $\Xi$'s in the maximal isospin channel, i.e., systems consisting only of 
neutrons and negative $\Xi$'s or protons and neutral $\Xi$'s, are bound.
As mentioned above, the uniqueness of these systems is a consequence
of the two-body interactions between $NN$, $\Xi N$ 
and $\Xi\Xi$ pairs being all in the isospin 1 channel. Thus,
the strong decay $\Xi N \to\Lambda\Lambda$ is forbidden. 
Therefore, such states, if bound, would be stable under the 
strong interaction. This is why we now proceed to study four-body
systems made of $N$'s and $\Xi$'s in the maximal isospin channel, $I=2$.
The most favorable configuration to minimize the effect of the Pauli
principle is the $\Xi\Xi NN$ system, that due to identity of two $N$'s and two
$\Xi$'s can only exist with $J=0$~\cite{Gai16}.

The binding energy of the $\Xi\Xi NN$ state has been calculated by means of the
variational method with generalized Gaussians described in Sec.~\ref{secII}. The
method has been used in the four-body sector to study the possible existence of 
tetraquarks~\cite{Vij07,Vin09,Car11} and tested against the hyperspherical harmonic 
formalism with comparable results~\cite{Vij09,Via09}. We show in Fig.~\ref{fig5} the
binding energy of the $(I)J^P=(2)0^+$ $\Xi\Xi NN$ state as a function of
the number of Gaussians in the variational calculation. As we can see the result 
is almost stable considering 12 Gaussians, although we have pushed further our calculation 
with a negligible gain of binding in the second decimal digit. The lowest threshold for this
state is $2B_2=$ 3.2 MeV, where $B_2$ is the binding energy of the $D^*$ $\Xi N$ state (see
Table~\ref{t2}). Thus, the state lies 7.4 MeV below the $\Xi\Xi NN$ mass, with a separation
energy of 4.2 MeV with respect to an asymptotic state made of two $D^*$ $\Xi N$ dibaryons.

One can also study the behavior of the root mean square radius (RMS) of the
four-body system, defined in the usual way,
\begin{eqnarray}
{\rm RMS} &=& \left({\frac{\sum_{i=1}^{4} m_i \br (\vec{r}_i-
\vec{R}_{CM})^2\et}{\sum_{i=1}^{4} m_i}} \right)^{1/2} \nonumber \\
& = & \frac{1}{2}\left( 
\frac{\br r_{NN}^2\et}{1+m_\Xi/m_N} + \br r_{\Xi\Xi}^2\et \frac{m_\Xi/m_N}{1+m_\Xi/m_N} +
\br r_{NN-\Xi\Xi}^2\et \frac{m_\Xi/m_N}{\left(1+m_\Xi/m_N\right)^2}
\right)^{1/2} \, . 
\end{eqnarray}
The results are shown in Fig.~\ref{fig6}, where besides the RMS radius we have also
calculated the root mean square radii of the different Jacobi coordinates. As seen in
Table~\ref{t2}, only the $^1S_0(I=1)$ $NN$ and $\Xi\Xi$ channels contribute to the
$(I)J^P=(2)0^+$ $\Xi\Xi NN$ state. As discussed in Sec.~\ref{secIV}, although
they are attractive, the $^1S_0(I=1)$ $NN$ and $\Xi\Xi$ channels do not present a 
bound state, giving the largest internal radii. In the $\Xi N$ subsystem one finds 
contributions from the $^1S_0(I=1)$ and $^3S_1(I=1)$ channels, the last one 
presenting the $D^*$ bound state, which is the responsible of the
smallest radius in the $\Xi - N$ relative coordinate. The RMS gets fully stabilized with
14 Gaussians with a value of 1.18 fm.

We have finally evaluated
the binding energy of the $\Lambda\Lambda NN$ system with quantum numbers
$(I)J^P=(1)0^+$~\cite{Gar17}. The system is unbound appearing just above threshold and thus
it does not seem to be Borromean, a four-body bound
state without two- or three-body stable subsystems. An unbound result was also
reported in Ref.~\cite{Lek14}, although in this case the authors made use
of repulsive gaussian-type potentials for any of the
two-body subsystems (see the figure on pag. 475) what does not allow
for the existence of any bound state. 
\begin{figure*}[t]
\vspace*{-.5cm}
\resizebox{9.cm}{13.cm}{\includegraphics{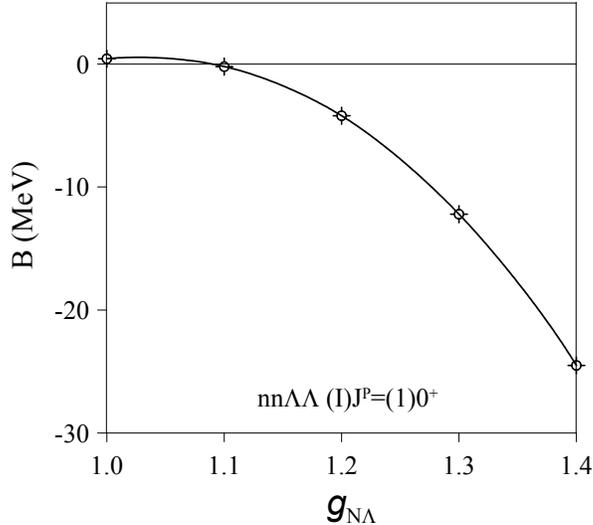}}
\vspace*{-5.0cm}
\caption{Binding energy of the $(I)J^P=(1)0^+$ $nn\Lambda\Lambda$ state as a function of
the multiplicative factor, $g_{N\Lambda}$, in the attractive part of $V^{N\Lambda}(r)$ interaction 
for $g_{NN}=g_{\Lambda\Lambda}=1$.}
\label{fig7}
\end{figure*}

We have studied the dependence of the binding on the strength of
the attractive part of the different two-body interactions entering 
the four-body problem. For this purpose we have used the following
interactions,
\begin{equation}
V^{B_1B_2}(r)=-g_{B_1B_2} \, A\,\frac{e^{-\mu_Ar}}{r}+B\, \frac{e^{-\mu_Br}}{r} \, 
\label{eq21new} 
\end{equation}
with the same parameters given in Table~\ref{t1}.
The system hardly gets bound for a reasonable increase of the strength of the 
the $\Lambda\Lambda$, $g_{\Lambda\Lambda}$, interaction. 
Although one cannot exclude that the genuine $\Lambda\Lambda$ interaction in 
dilute states as the one studied here could be slightly stronger that the one 
reported in Ref.~\cite{Nag15}, however, one needs $g_{\Lambda\Lambda} \ge 1.8$ to
get a bound state, what would destroy the agreement with the ESC08c Nijmegen 
$\Lambda\Lambda$ phase shifts. Note also that this is a very sensitive 
parameter for the study of double-$\Lambda$ hypernuclei~\cite{Nem03}
and this modification would produce an almost $\Lambda\Lambda$ bound state
in free space, in particular it would give rise to $a_{^1S_0}^{\Lambda\Lambda} = -29.15$ fm
and ${r_0}_{^1S_0}^{\Lambda\Lambda} = 1.90$ fm. 
The four-body system would also become bound taking a factor $1.2$ in the $NN$ interaction.
However, such modification would make the $^1S_0$ $NN$ potential as strong as the $^3S_1$~\cite{Mal69} and 
thus the singlet $S-$wave would develop a dineutron bound state, $a_{^1S_0}^{NN} = 6.07$ fm
and ${r_0}_{^1S_0}^{NN} = 1.96$ fm.
The situation is slightly different when dealing with the $\Lambda N$ interaction. 
We have used a common factor $g_{N\Lambda}$ for attractive part of the two $\Lambda N$
partial waves, $^1S_0$ and $^3S_1$. We show in Fig.~\ref{fig7} 
the binding energy of the $(I)J^P=(1)0^+$ $\Lambda\Lambda NN$ state as a function of
the multiplicative factor $g_{N\Lambda}$, for $g_{NN}=g_{\Lambda\Lambda}=1$.
As one can see the four-body system develops a bound state for $g_{N\Lambda}=1.1$,
giving rise to the $\Lambda N$ low-energy parameters: $a_{^1S_0}^{\Lambda N} = -5.60$ fm,
${r_0}_{^1S_0}^{\Lambda N} = 2.88$ fm, $a_{^3S_1}^{\Lambda N} = -2.91$ fm, and
${r_0}_{^3S_1}^{\Lambda N} = 2.99$ fm, far from the values constrained by the
existing experimental data.

Ref.~\cite{Ric15} tackled the same problem by fitting low-energy parameters
of older versions of the
Nijmegen-RIKEN potential~\cite{Rij10,Rij13} or chiral effective field 
theory~\cite{Pol07,Hai13}, by means of
a single Yukawa attractive term or a Morse parametrization. 
The method used to solve the four-body problem is similar to the one
we have used in our calculation, thus the results might be directly 
comparable. Our improved description of the two- and three-body subsystems
and the introduction of the repulsive barrier for the $^1S_0$ $NN$ partial wave,
relevant for the study of the triton binding energy (see Table II 
of Ref.~\cite{Mal70}), leads to a four-body state above threshold,
that cannot get bound by a reliable modification in the
two-body subsystems. As clearly explained in Ref.~\cite{Ric15}, the window of
Borromean binding is more an more reduced for potentials 
with harder inner cores. 

\section{Summary}
\label{secVI}

This manuscript intends to summarize our recent work on few-body systems
made of $N$'s, $\Lambda$'s and $\Xi$'s based on the most recent updates of the 
ESC08c Nijmegen potential in the different
strangeness sectors, accounting for the recent experimental information. 
We have solved the three- and four-body bound state problems by means of Faddeev 
equations and a generalized Gaussian variational method, respectively.
The hypertriton, $np\Lambda$ $(I)J^P=(1/2)1/2^+$, is bound by 144 keV, and
the recently discussed $nn\Lambda$ $(I)J^P=(1/2)1/2^+$ system is unbound.
We have found that the $\Xi NN$ system presents bound states with quantum numbers
$(I)J^P=(3/2)1/2^+$ and $(1/2)3/2^+$, the last one being a deeply bound state lying 
15 MeV below the $\Xi d$ threshold. 
The $\Xi\Xi N$ system presents a bound state with quantum numbers $(I)J^P=(3/2)1/2^+$,
in spite of having used the most recent update of the ESC08c Nijmegen potential that does
not predict $\Xi\Xi$ bound states.
In the case of the three-body systems we note that there appear bound states
in all systems made of $N$'s and $\Xi$'s with maximal isospin.
The same conclusion has been obtained in the four-body system, concluding a
$\Xi\Xi NN$ bound state with quantum numbers $(I)J^P=(2)0^+$,
lying 7.4 MeV below the $\Xi\Xi NN$ threshold with a root mean square radius 
of 1.18 fm. We have also studied the $(I)J^P=(1)0^+$ $\Lambda\Lambda NN$
state, it does not present a bound state. Thus, the ${}_{\Lambda\Lambda}^{\,\,\,\,4}n$ four-body system does 
not seem to be Borromean.

\section{acknowledgments} 

This work has been partially funded by COFAA-IPN (M\'exico), 
by Ministerio de Econom\'\i a, Industria y Competitividad 
and EU FEDER under Contracts No. FPA2016-77177 and FPA2015-69714-REDT,
by Junta de Castilla y Le\'on under Contract No. SA041U16,
by Generalitat Valenciana PrometeoII/2014/066,
and by USAL-FAPESP grant 2015/50326-5.

\end{document}